# EXISTENCE OF A NEW QUANTUM PHASE IN EXACTLY SOLVABLE ANTIFERROMAGNETIC ISING-HEISENBERG MODELS ON PLANAR LATTICES

Michal JAŠČUR and Jozef STREČKA

Department of Theoretical Physics and Astrophysics, Institute of Physics,
P.J. Šafárik University, Moyzesova 16, 041 54 Košice, Slovak Republic,
E-mail: jascur@kosice.upjs.sk, jozkos@pobox.sk

**SUMMARY**

In this work we deal with doubly decorated Ising-Heisenberg models on planar lattices. Applying the generalized decoration-iteration transformation we obtain exact results for the antiferromagnetic version of the model. The existence of a new quantum dimerized phase is predicted and its physical properties are studied and analyzed. Particular attention has been paid to the investigation of the phase boundaries, pair-correlation functions and specific heat. A possible application of the present work to some molecular magnets is also drawn.

## 1. INTRODUCTION

The low-dimensional magnetic materials have attracted considerable experimental and theoretical interest since Haldane's pioneering work [1] that pointed out the difference between the integer spin Heisenberg quantum antiferromagnets (QHA) and half integer ones. Among the most fascinating problems being of current research interest in this field one should mention: the quantum phase transitions [2], spin-Peierls instabilities [3], quantum entanglement [4], magnetization plateau [5] and so on. In addition to the above mentioned topics, the two-dimensional spin-1/2 QHA have been also very intensively studied in connection with possible applications to the layered perovskites and cooper-oxide compounds. In fact, in the insulating cuprates appears Cu-O-Cu superexchange coupling that is responsible for the antiferromagnetic ordering of the cooper spins in the $CuO_2$ layers. The existence of the superconducting phase is then associated with quantum fluctuations [6] or with frustrations of the exchange interactions caused by doping. [7].

It is worth noticing that despite of extensive studies only few exact results have been obtained in this research field so far [8]. This principal mathematical intractability is closely related to the many-body effects, as well as, to the non-commutability of the spin operators entering the relevant Heisenberg Hamiltonian. The main purpose of this work is to introduce a wide class of exactly solvable antiferromagnetic two-dimensional models that enable to study interesting quantum phenomena. A possible application of the presented models to some realistic molecular-based magnets will be also drawn. The paper is organized as follows. In Sec.2 we briefly review the main points of the applied method. The numerical results for the ground state, phase diagrams, pair-correlation functions and specific heat are discussed in Sec. 3 and concluding remarks are summarized in Sec. 4.

## 2. FORMULATION

In this work we will study a spin-1/2 doubly decorated planar Ising-Heisenberg models described by the Hamiltonian

$$H = \sum_{i,j} J\left[\Delta\left(S_i^x S_j^x + S_i^y S_j^y\right) + S_i^z S_j^z\right] + \sum_{k,l} J_1 S_k^z \mu_l^z, \quad (1)$$

where $S_k^\alpha (\alpha = x, y, z)$ and $\mu_l^z$ denote the components of the standard spin–1/2 operators and the summations are carried out over the nearest neighbors only. The exchange parameters $J$ and $J_1$ couple two Heisenberg atoms or neighboring pairs consisting of Ising and Heisenberg atoms, respectively. Moreover, we will assume that both interactions are positive i.e. supporting the antiferromagnetic ordering. Finally, the parameter $\Delta$ reflects the anisotropy in the interaction between Heisenberg atoms. In fact, by changing the strength of $\Delta$ one can control the behaviour of the system between the Ising-regime (or the easy-axis anisotropy ($\Delta<1$)) and the XY-regime (or the easy-plane anisotropy ($\Delta>1$)).

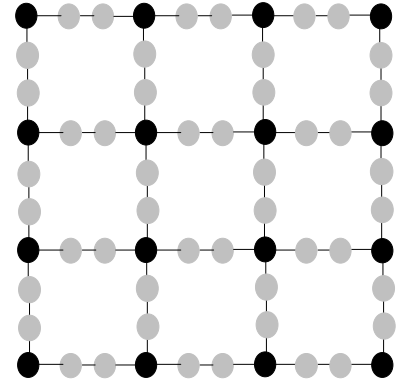

**Fig. 1** Doubly decorated Ising-Heisenberg model on the square lattice. Black circles denote the Ising atoms and the gray the Heisenberg ones.

As usual, the quantity of our primary interest is the magnetic partition function defined by

$$Z = \text{Tr} \exp(-\beta H). \tag{2}$$

The basic idea of our calculation is to find a relation between the partition function of our system and that of the standard spin-1/2 Ising model on the relevant undecorated lattice. For this purpose, we rewrite the Hamiltonian (1) in terms of the site Hamiltonians. Namely,

$$H = \sum_k H_k, \tag{3}$$

with

$$H_k = J\left[\Delta(S_{k1}^x S_{k2}^x + S_{k1}^y S_{k2}^y) + S_{k1}^z S_{k2}^z\right] + J_1(S_{k1}^z \mu_{k1}^z + S_{k2}^z \mu_{k2}^z). \tag{4}$$

Now, substituting this relation into Eq. (2) and taking into account the commutability of the site Hamiltonians ($[H_k, H_l]=0$ if $k \neq l$), one can easily express the partition function in the form

$$Z = \text{Tr} \exp(-\beta H) = \text{Tr} \exp\left(-\beta \sum_{k=1}^{Nq/2} H_k\right) = \text{Tr}_{\{\mu\}} \prod_{k=1}^{Nq/2} \text{Tr}_{S_{k1}} \text{Tr}_{S_{k2}} \exp(-\beta H_k), \tag{5}$$

where $\beta=1/k_B T$ and $q$ is the coordination number of the undecorated lattice and $N$ represents the total number of atoms on the original lattice. The symbol $\text{Tr}_{\{\mu\}}$ denotes the trace over all Ising spin variables and $\text{Tr}_{S_{k1}} \text{Tr}_{S_{k2}}$ means the trace over a couple of the Heisenberg atoms on the $k$th bond.

In order to proceed further, we introduce the following generalized decoration-iteration transformation

$$\text{Tr}_{S_{k1}} \text{Tr}_{S_{k2}} \exp(-\beta H_k) = 2\exp(-\beta J/4)\cosh\left[\beta J_1(\mu_{k1}^z + \mu_{k2}^z)/2\right] + 2\exp(\beta J/4)\cosh\left[\beta \sqrt{J_1^2(\mu_{k1}^z - \mu_{k2}^z)^2 + (J\Delta)^2}/2\right] = A\exp(\beta R \mu_{k1}^z \mu_{k2}^z), \tag{6}$$

that enables one to rewrite equation (5) as follows

$$Z(\beta, J, J_1, \Delta) = A^{Nq/2} Z_0(\beta, R). \tag{7}$$

Here, $Z_0$ represents the partition function of the standard spin-1/2 Ising model on the relevant undecorated lattice and the unknown parameters $A$ and $R$ can be straightforwardly expressed as a function of $\beta$, $J$, $J_1$ and $\Delta$. Namely,

$$A = 2\exp(-\beta J/2)(W_1 W_2)^{1/2},$$
$$\beta R = 2\ln\left(\frac{W_1}{W_2}\right). \tag{8}$$

where we have introduced the functions $W_1$ and $W_2$

$$W_1 = \cosh(\beta J_1/2) + \exp(\beta J/2)\cosh(\beta J \Delta/2),$$
$$W_2 = 1 + \exp(\beta J/2)\cosh\left(\beta \sqrt{J_1^2 + (J\Delta)^2}/2\right). \tag{9}$$

Equations (7)-(9) represent a complete mathematical correspondence between the model under investigation and the relevant exactly solvable spin-1/2 Ising model on the original (undecorated) lattice. It is worthy to note that from equations (7)-(9) one can, in principle, derive relations for all physical quantities of interest applying standard thermodynamic equations. However, this approach leads in practice to very lengthy and tedious calculations. Fortunately, this problem can be avoided using the following exact equations

$$\langle f_1(\mu_i^z, \mu_j^z, ..., \mu_k^z)\rangle_d = \langle f_1(\mu_i^z, \mu_j^z, ..., \mu_k^z)\rangle_0,$$
$$\langle f_2(S_{k1}^\alpha, S_{k2}^\gamma, \mu_{k1j}^z, \mu_{k2}^z)\rangle_d = \left\langle \frac{\text{Tr}_{S_{k1}}\text{Tr}_{S_{k2}} f_2(S_{k1}^\alpha, S_{k2}^\gamma, \mu_{k1j}^z, \mu_{k2}^z)\exp(-\beta H_k)}{\text{Tr}_{S_{k1}}\text{Tr}_{S_{k2}} \exp(-\beta H_k)}\right\rangle_0, \tag{10}$$

where $f_1$ represents a function depending exclusively on Ising-spin variables and $f_2$ denotes a function which depends on the spins operators of the $k$th bond. The superscripts $\alpha, \gamma \equiv (x, y, z)$ denote the components of the spin operators and finally the symbol $\langle ...\rangle_d$ and $\langle ...\rangle_0$ represent the standard ensemble average related to the decorated and original model, respectively.

Now, applying one of the standard methods (see for example [9]), one easily derives from equations (10) relatively simple expressions for the pair-correlation functions. Namely,

$$q_{ii}^{zz} \equiv \langle \mu_{k1}^z \mu_{k2}^z\rangle_d = \langle \mu_{k1}^z \mu_{k2}^z\rangle_0,$$
$$q_{hh}^{xx} \equiv \langle S_{k1}^x S_{k2}^x\rangle_d \equiv \langle S_{k1}^y S_{k2}^y\rangle_d =$$
$$= -(K_1 + K_2)/8 - q_{ii}^{zz}(K_1 - K_2)/2,$$

$$q_{hh}^{zz} \equiv \langle S_{k1}^z S_{k2}^z \rangle_d = 1/4 - (K_3 + K_4)/4 - q_{ii}^{zz}(K_3 - K_4),$$

$$q_{ih}^{zz} \equiv \langle (S_{k1}^z \mu_{k1}^z + S_{k2}^z \mu_{k2}^z)/2 \rangle_d =$$
$$= -(K_5 + K_6)/8 - q_{ii}^{zz}(K_5 - K_6)/2, \quad (11)$$

where the superscripts specify the space-direction, the subscripts identify the type of the atoms and $\langle \mu_{k1}^z \mu_{k2}^z \rangle_0$ represents the nearest-neighbor correlation on the relevant undecorated lattice which is well known. The coefficients in equation (11) depend on the parameters of the Hamiltonian and temperature, and they are given by

$$K_1 = \exp(\beta J/2)\sinh(\beta J\Delta/2)/W_1,$$

$$K_2 = \frac{(J\Delta)\exp(\beta J/2)}{W_2\sqrt{J_1^2 + (J\Delta)^2}} \sinh\left(\beta\sqrt{J_1^2 + (J\Delta)^2}/2\right),$$

$$K_3 = \exp(\beta J/2)\cosh(\beta J\Delta/2)/W_1,$$

$$K_4 = \exp(\beta J/2)\cosh\left(\beta\sqrt{J_1^2 + (J\Delta)^2}/2\right)/W_2,$$

$$K_5 = \sinh(\beta J_1/2)/W_1, \quad (12)$$

$$K_6 = \frac{J_1 \exp(\beta J/2)}{W_2\sqrt{J_1^2 + (J\Delta)^2}} \sinh\left(\beta\sqrt{J_1^2 + (J\Delta)^2}/2\right).$$

With the help of the correlation functions we can express the internal energy of the system in the form:

$$U = N q J \Delta\, q_{hh}^{xx} + N q J\, q_{hh}^{zz}/2 + N q J_1\, q_{ih}^{zz}, \quad (13)$$

and then we can easily also calculate the specific heat from the relation $C = \left(\dfrac{\partial U}{\partial T}\right)$.

## 3. NUMERICAL RESULTS AND DISCUSSION

In this part we will discuss the most interesting findings for the Ising-Heisenberg model on the doubly decorated square lattice which entirely illustrates the behaviour of the considered doubly decorated models.

In order to understand the ground-state properties of the system, we have analyzed the relevant pair-correlation functions that are for $T = 0$ given by

$$q_{hh}^{zz} = -0.25, \quad q_{hh}^{xx} = -\frac{J\Delta}{4\sqrt{J_1^2 + (J\Delta)^2}}, \quad (14)$$

$$q_{ii}^{zz} = -0.25, \quad q_{ih}^{zz} = -\frac{J_1}{4\sqrt{J_1^2 + (J\Delta)^2}}. \quad (15)$$

As one can expect, we have found that the pairs consisting of two Heisenberg atoms, as well as, those consisting of two Ising atoms align antiparalel (i. e. the relevant correlation functions take the maximum possible value). Surprisingly, excepting this trivial observation one also finds that the correlation between Ising and Heisenberg atoms does not take its maximum value ($q_{ih}^{zz} \neq -0.25$) for $\Delta \neq 0$. Moreover, there always exists a short-range order in the $xy$-plane since $q_{hh}^{xx} = q_{hh}^{yy} \neq 0$. These interesting results are depicted in Fig. 2 for the case of $J_1/J = 1.0$. In fact, such a behavior of the system is closely related to the Heisenberg uncertainty principle and it leads to the existence of an unusual quantum phase that will be called in this work as a quantum dimerized phase. It follows from our calculation that this phase is the only stable phase at the ground state and it can be characterized as the quantum phase in which two different contributions are combined. The first one is originated by the spins of the Ising sublattice exhibiting perfect antiferromagnetic order and the second one coming from the Heiseberg atoms that apparently create antiferromagnetic dimers on the bonds of the original lattice. However, the most fascinating finding is the fact that the above-mentioned dimers are oriented randomly with respect to their Ising nearest neighbors. Consequently, the quantum dimerized phase is partly disordered and it may exhibit some different features in comparison with the ordinary long-range ordered phases.

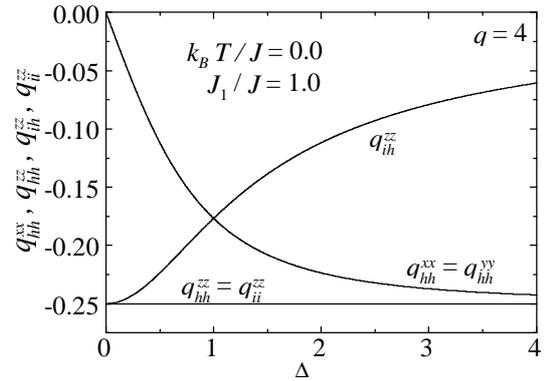

**Fig. 2** The ground-state pair-correlation functions versus anisotropy

Now, let us proceed to investigate the behavior of the system at finite temperatures. Before discussing numerical results, it is useful to note that the phase diagrams of our decorated system can by simply calculated after substituting the critical temperature of the original lattice ($\beta_c R = 2\ln(1+\sqrt{2})$) into equation (8).

At first, we have depicted in Fig. 3 the phase boundaries in the $\Delta - T$ plane for different values of the exchange parameter $J_1$. As one can see, in the case of the weak exchange parameter $J_1$, the critical temperature monotonically decreases from its pure Ising limit at $\Delta = 0$ and tends to zero for $\Delta \to \infty$. On the other hand, if the exchange parameter $J_1$ becomes strong enough, the critical temperature at first exhibits a broad maximum and then again goes to zero (see the case $J_1/J = 3.0$). These characteristic dependences appear due to the competitive influences of the exchange interactions $J, J_1$, anisotropy parameter $\Delta$ and temperature. In order to understand more deeply the nature of the phase transition between the dimerized and disordered phase, we have studied the temperature variation of the pair-correlation functions, as well as, the thermal dependences of the specific heat.

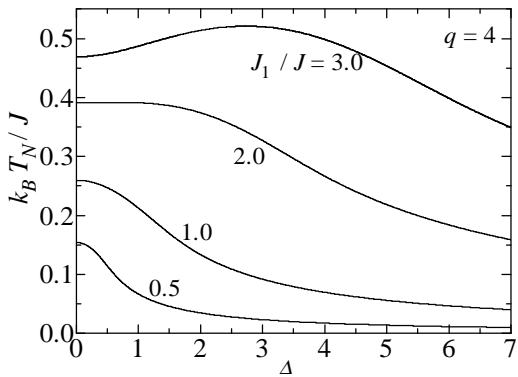

**Fig.3** The phase diagram in the $\Delta - k_B T_N / J$ plane.

In Fig. 4, there are presented the temperature dependences of the pair-correlation functions between two Heisenberg-type atoms in the $z$-direction (dashed lines) together with those in the $x$-direction (solid lines) for some typical sets of parameters. Moreover, in Fig. 5 we show the same dependences (but only in the $z$-direction) also for the pairs of Ising spins and those consisting of one Ising and one Heisenberg atoms. As one can see, in agreement with the ground state analysis the correlation functions $q_{ii}^{zz}$ and $q_{hh}^{zz}$ take at $T = 0$ always the maximum possible value (–0.25) and then monotonically decrease with the temperature. Contrary to this standard behavior, all other correlation function exhibit non-trivial temperature dependences. Moreover, the detailed analysis around the phase-transition point reveals that all the correlation functions exhibit a weak energy-type singularity at $T = T_N$. This behavior indicates that the relevant phase transition can be of the second order. Indeed, the temperature dependences of the specific heat presented in Fig. 6 confirm this expectation. Moreover, one can verify that the specific heat has a logarithmic Onsager-type singularity at $T = T_N$. Thus one can conclude that the phase transitions between quantum dimerized and disordered phase belongs to the same universality class as that of the usual spin-1/2 2D Ising model. This can also be concluded from equations (7)-(9), from which one can see that the expression for the parameter $A$ is represented by analytical functions that cannot lead to new singularities.

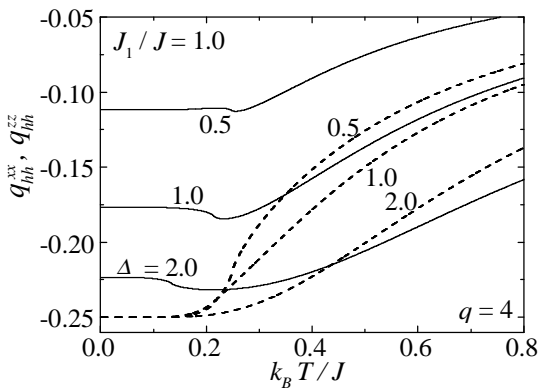

**Fig.4** The temperature variations of Heisenberg nearest–neighbor pair–correlation functions. The dashed (solid) lines corresponding to correlation in the $z$-direction ($x$-direction), respectively.

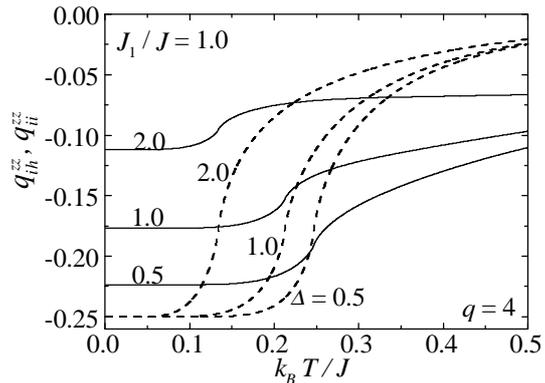

**Fig.5** The temperature variation of the Ising-Ising (dashed lines) and Ising-Heisenberg (solid lines) pair–correlation functions.

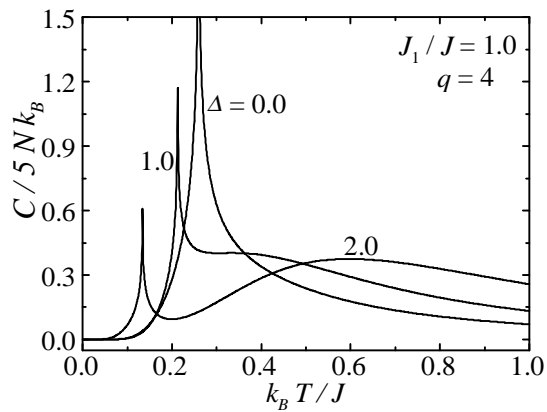

**Fig.6** The specific heat against temperature when exchange anisotropy is changed.

## 4. CONCLUSION

In this work we have investigated magnetic properties of the doubly decorated Ising-Heisenberg model. Applying the extended decoration-iteration transformation, we have exactly calculated the critical temperature, partition function, pair-correlation functions, internal energy and specific heat of the model. The most outstanding finding coming from this work is the prediction of the unusual quantum dimerized phase existing in the system in the whole range of parameters. This phase appears also in the ferromagnetic version of this model, however, it can exist only in the restricted region of the parameters [10].

Although, the numerical result have been presented only for the square lattice, it is clear that our conclusions can be simply generalized for many other planar Ising-Heisenberg models. Of course, the method developed in this work can by straightforwardly used to study doubly decorated Ising-Heisenberg linear chains. However, in this case the results will be completely different, since the spin-1/2 Ising linear chain has no phase transition at finite temperatures. Moreover, despite of the fact that the pure 3D Ising model has not been exactly solved, we can apply the present method to the 3D models, as well. In fact, we can obtain very accurate results for the phase boundaries if we combine the present formalism with the results known from the series-expansion methods.

On the other hand, our preliminary calculation indicates that the behavior of the doubly decorated Ising-Hesenberg models is significantly changed in arbitrary space-dimension if the Heisenberg atoms have spin 1.

Finally, we would like to emphasize that our theoretical predictions can by very useful in understanding of the magnetic properties of some real materials. The most promising from this point of view seem to be molecular-based magnets that have been recently synthesized by some authors [11]. These materials have the same structure as the model under investigation, therefore we hope that the existence of the quantum dimerized phase can be experimentally confirmed in the future.

*Acknowledgment:* This work has been supported by the Ministry of Education of Slovak Republic under VEGA grant No. 1/9034/02.

## BIOGRAPHY

Michal Jaščur was born on 16.10.1963. In 1987 he graduated (RNDr.) with distinction at the Department of Theoretical Physics and Geophysics of P. J. Šafárik University in Košice. He defended his PhD at alma mater in 1995. From 2000 he holds the position of the Associated Professor.

Jozef Strečka was born on 23.4.1977. In 2000 he graduated with distinction at the Department of Theoretical Physics and Geophysics of P. J. Šafárik University in Košice. At present, he is working as a PhD student at alma mater under the supervision of Associate Professor Michal Jaščur.

The research interests of both authors are related to the quantum theory of magnetism and phase transition phenomena.